# Spontaneous Spin Coherence in n-GaAs Produced by Ferromagnetic Proximity Polarization

R. J. Epstein, I. Malajovich, R. K. Kawakami, Y. Chye, M. Hanson,
P. M. Petroff, A. C. Gossard, and D. D. Awschalom

*Center for Spintronics and Quantum Computation, University of California, Santa Barbara, California 93106*

**Abstract**

We find that photoexcited electrons in an n-GaAs epilayer rapidly (< 50 ps) spin-polarize due to the proximity of an epitaxial ferromagnetic metal. Comparison between MnAs/GaAs and Fe/GaAs structures reveals that this coherent spin polarization is aligned antiparallel and parallel to their magnetizations, respectively. In addition, the GaAs nuclear spins are dynamically polarized with a sign determined by the spontaneous electron spin orientation. In Fe/GaAs, competition between nuclear hyperfine and applied magnetic fields results in complete quenching of electron spin precession.

Established methods of generating electron spin polarization in semiconductors include electrical spin injection and optical excitation. Electrical spin injection has been reported from magnetic semiconductors [1] and ferromagnetic metals. [2] Alternatively, optical excitation has been shown to induce coherent spin polarization in semiconductors that persists for ~0.1 μs,[3] and is robust to transport over distances of ~100 μm and across heterointerfaces. [4,5] In more recent work, it was shown that Mn-based hybrid ferromagnet/n-GaAs structures exhibit a ferromagnetic imprinting of the GaAs nuclear spin system.[6] The resulting nuclear polarization was found to track the magnetization of the ferromagnetic layer and influence the coherent spin dynamics of photoexcited electrons via the hyperfine interaction. Furthermore, initial evidence was provided suggesting that the ferromagnet also generates coherent electron spin polarization in the GaAs, possibly driving the dynamic polarization of nuclear spin.

Here, we observe spontaneous electron spin coherence resulting from this ferromagnetic proximity polarization (FPP) process in similar hybrid structures, including Fe/GaAs. Time-resolved Faraday rotation [7] (TRFR) measurements reveal that photoexcited electrons rapidly spin-polarize (< 50 ps) in the presence of the ferromagnetic layer and maintain spin coherence for several nanoseconds. Comparison between MnAs/GaAs and Fe/GaAs structures indicates that these spins align antiparallel or parallel to the ferromagnet's magnetization **M**, respectively. In addition to restricting the possible mechanisms underlying FPP, this opposing alignment reveals that the sign of the nuclear spin polarization depends on the FPP electron spin orientation rather than **M**. One manifestation of this dependence is the quenching of coherent electron spin precession in an applied magnetic field (of order 1 T) in Fe/GaAs. These measurements indicate that FPP electrons cause the dynamic polarization of nuclear spin. Moreover, they demonstrate an additional avenue of control over both electron and nuclear spins in a semiconductor through the choice of ferromagnetic material and the orientation of **M**.

We investigate a series of hybrid ferromagnet/semiconductor samples grown by molecular beam expitaxy with the following structure: ferromagnet/n-GaAs(100 nm)/Al$_{0.75}$Ga$_{0.25}$As(400 nm)/n$^+$-GaAs(001)-substrate, where the ferromagnet is either 25 nm of MnAs (type A)[8] or 10 nm of Fe.[9] The Fe is deposited at room temperature and capped with 50 Å of Al to prevent oxidation, and the n-type doping (Si: ~7 x 10$^{16}$ cm$^{-3}$) of the GaAs layer is chosen to extend the electron spin lifetime.[3] A control sample is also grown with the above structure but without the ferromagnetic layer. Samples are mounted on fused silica and the GaAs substrate is removed by a chemically selective etch. The magnetic properties of these thin films are characterized at $T$ = 10 K with a superconducting quantum interference device (SQUID) magnetometer.

Electron spin dynamics in these samples are measured with TRFR using a mode-locked Ti:Sapphire laser that emits ~100 fs pulses at a 76-MHz repetition rate. The output, with energy tuned near the GaAs band gap (~1.52 eV at $T$ = 5 K), is split into circularly polarized (CP) pump and linearly polarized (LP) probe beams that are focused to overlapping ~100 μm spots on the sample. Due to optical orientation governed by the selection rules in GaAs,[10] a given pump pulse excites carriers with net spin aligned along the pump path. After a time $\Delta t$, the corresponding probe pulse passes through the sample and its polarization axis rotates by an angle $\theta$ proportional to the projection of the net spin onto the probe path. An external field **B** is applied transverse to the pump beam, inducing coherent spin precession in the plane perpendicular to the field. This is manifested as a periodic oscillation of $\theta$ vs $\Delta t$, where $\Delta t$ is varied with a mechanical delay line. Alternatively, separate lasers can be used for the pump and probe, enabling independent tuning of their energies. In this two-color TRFR, $\Delta t$ is varied by actively synchronizing the two pulse trains.[11] Finally, we also use a LP pump (which generates unpolarized carriers) to investigate polarization processes distinct from optical orientation.

Figure 1(left inset) shows the experimental configuration, where the pump beam is normal to the MnAs/GaAs surface and the in-plane magnetization is roughly perpendicular to both **B** and the optical path. The solid symbols in Fig. 1 represent a TRFR scan taken with a CP pump at $B = 0.12$ T. In this case, the net spin along **x**, $S_x^{CP}(\Delta t)$, exhibits the oscillatory behavior discussed above. In contrast, while a LP pump does not generate spin polarization in the control sample (as expected from the selection rules),[10] samples with a MnAs layer exhibit an oscillatory signal, $S_x^{LP}(\Delta t)$, similar to $S_x^{CP}(\Delta t)$ but with a ~90° phase shift to the precession (Fig. 1, open circles).[12] Thus, the ferromagnetic MnAs layer appears to generate a spontaneous spin polarization in the GaAs (the FPP effect).

For either pump polarization, $S_x(\Delta t)$ is described by

$$S_x(\Delta t) = S_0 e^{-\Delta t/T_2^*} \cos(\omega \Delta t + \phi), \qquad (1)$$

where $S_0$ is the amplitude, $T_2^*$ is the effective transverse spin lifetime, $\omega$ is the Larmor precession frequency, and $\phi$ is the phase of the spin precession. Here, $\omega = g\mu_B B_{tot}/\hbar$, where $g$ is the electron g-factor, $\mu_B$ is the Bohr magneton and $B_{tot}$ is the total magnetic field, which includes $B$ and any local fields. Several aspects of the FPP process can be quantified by fitting the data in Fig. 1. First, the energy dependence of $S_x(\Delta t)$ is measured with two-color TRFR by varying the pump energy $E_{pump}$, while holding the probe energy constant. For both CP and LP pumps, a plot of $S_0$ vs $E_{pump}$ (Fig. 1, inset), where $S_0$ has been normalized for ease of comparison, shows that the onset of TRFR with increasing energy coincides with the GaAs absorption edge. Second, calculating the g-factor from the fitted $\omega$ yields $g = -0.4$,[13] which is consistent with the value of $g = -0.44$ reported for nominally undoped GaAs.[14] These results confirm that the measured electron spins reside in the GaAs layer.

In addition, we find that $S_0^{LP} \sim 1/3\, S_0^{CP}$, indicating that the efficiency of FPP is comparable to optical orientation. Furthermore, it is assumed in Eq. (1) that there is no rise time associated with $\mathbf{S}^{LP}$. Therefore, the negligible deviation at small $\Delta t$ of the fit (Fig.1, solid line) from the data sets an upper bound on such a rise time of ~50 ps. While this shows that the spontaneous polarization of spin is rapid, spin coherence times remain quite long. We obtain $T_2^* \sim 4$ ns for a LP pump, whereas $T_2^* \sim 2$ ns for a CP pump.[12] Finally, the initial direction of $\mathbf{S}^{LP}$ can be inferred from $\phi$. For $S_x^{LP}(\Delta t)$, we obtain $\phi \sim -91°$, which is consistent with spin initially polarized antiparallel to $\mathbf{M}$ (in-plane).

We note that these measurements of FPP rely on the strong, uniaxial, in-plane magnetic anisotropy of the MnAs.[8] Hysteresis curves of the (epitaxial) MnAs/GaAs show that the [110] (GaAs) crystal direction is magnetically easy, characterized by sharp switching at the coercive field $B_c = \pm 0.086$ T. The $[\bar{1}10]$ axis is magnetically hard, exhibiting negligible remanence and saturation above ~2.5 T. Due to this strong anisotropy, $\mathbf{M}$ is effectively pinned along [110] at low fields. In the measurements above, we set the angle $\alpha$ between [110] and $\mathbf{B}$ to ~86° so that $\mathbf{M}$ is roughly perpendicular to $\mathbf{B}$ (applied in-plane) and switches at ~1.2 T = $B_c/\cos(86°)$. Spin that is aligned along **M** will then have a large component perpendicular to **B** that precesses and is detectable by TRFR.

Making further use of this anisotropy, we verify that $\mathbf{S}^{LP}$ is controlled by **M** by measuring the field dependence of $S_x^{LP}(\Delta t)$. Figure 2(a) shows a gray-scale plot of $S_x^{LP}(\Delta t)$ vs $B$ in the same geometry as Fig. 1. As $B$ is swept from -3 T to 3 T, the TRFR oscillations change sign twice. The first sign change occurs as $B$ crosses zero in accordance with Eq. (1), which is an odd function of $\omega$ when $\phi = -90°$. The second sign change, at ~1.2 T, is due to the reversal of **M** along the MnAs easy axis as discussed above. Fitted values of $S_0^{LP}$ as a function of $B$ are also plotted (solid circles), and a comparison with the opposite sweep direction (open circles) confirms that $S_0^{LP}$ follows the magnetization hysteresis.

In contrast, data taken with a CP pump [Fig. 2(b)] do not clearly exhibit the hysteresis seen in $S_x^{LP}(\Delta t)$. For clarity, let us denote FPP spin as $\mathbf{S}^{FPP}$ for either pump polarization. Assuming that the FPP contribution to $S_x^{CP}(\Delta t)$ is comparable to that of $S_x^{LP}(\Delta t)$, reversing the sign of $\mathbf{S}^{FPP}$ (upon crossing zero field or switching **M**) is expected to result in a ~40° phase shift in $S_x^{CP}(\Delta t)$. However, we do not observe phase shifts of this magnitude,[15] indicating that TRFR due to $\mathbf{S}^{FPP}$ depends on the pump polarization (LP or CP).

We now verify that $\mathbf{S}^{FPP}$ aligns along the magnetization axis by considering the dependence of $S_x^{LP}(\Delta t)$ on the orientation of **M**. In Fig. 2(c), a decrease in $S_0^{LP}$ is observed when the sample is rotated about its normal, reducing the angle α between the magnetic easy axis and **B** (~0.12 T). In this case, $|S_0^{LP}|$ (solid squares) falls off as $\sin(\alpha)$ (solid line), tracking the component of **M** perpendicular to **B**. In addition, we note a reduction of $|S_0^{LP}|$ with increasing $|B|$ in Fig. 2(a), as **M** is rotated towards **B** in accordance with the magnetic properties of the MnAs. Both of these results are consistent with **S**$^{FPP}$ aligned (anti-) parallel to **M**.

In an attempt to gain insight into the microscopic origin of the FPP effect, we compare MnAs/GaAs and Fe/GaAs samples. In contrast to MnAs/GaAs, Fe/GaAs exhibits weaker in-plane magnetic anisotropy. Magnetization measurements reveal significant remanence along [$\bar{1}10$], indicating that **M** is not pinned along [110]. However, out-of-plane hysteresis data show that both thin film materials have negligible remanence and high saturation fields (> 3 T) along [001]. Exploiting this similarity, we compare the two samples by setting $\alpha = 0°$, rotating the samples 30° about the [$\bar{1}10$] axis [Figs. 3(a) and (b)], and raising the temperature to 110 K to remove the possible influence of nuclear polarization.[6] The out-of-plane anisotropy then prevents **M** from aligning with **B** (for $|$**B**$| < 3$ T). In this orientation, refraction results in an optical path that is no longer perpendicular to **B**. Consequently, the component of spin along **B** also contributes to TRFR as a non-oscillatory signal. A term of the form $S_{\parallel}\exp(-\Delta t/T_1)$ is therefore added to Eq. (1) upon fitting the following data, where $T_1$ is the longitudinal spin lifetime, and $S_{\parallel}$ is the amplitude.[16]

Figures 3(c) and (d) display plots of $S_x^{CP}(\Delta t)$ (open circles) at $B = +0.25$ T for MnAs/GaAs and Fe/GaAs, respectively. The data are nearly identical for the two samples and are insensitive to the sign of $B$. In contrast, Figs. 3(e) and (f) show plots of $S_x^{LP}(\Delta t)$ at $B = \pm 0.25$ T for MnAs/GaAs and Fe/GaAs, respectively. While the TRFR amplitudes from both samples have similar magnitudes, they are of opposite sign for a given field. By comparing the signs and phases obtained from the fits with those of $S_x^{CP}(\Delta t)$, the projections onto **M** of the initial spin vectors are inferred: Fe polarizes electron spins in GaAs parallel to **M** whereas MnAs polarizes them antiparallel.

These observations restrict the possible origins of FPP. For instance, magnetic fringe fields cannot explain the opposite polarizations for Fe and MnAs. Further, magnetic circular dichroism effects should be minimal when **M** is perpendicular to the optical path and optical orientation of electron spin along **M** is inefficient in this case assuming bulk selection rules.[10] One possible explanation of FPP is a spin interaction at the ferromagnet/GaAs interface governed by the spin-dependent density of states (DOS).[17] Experimental evidence suggests that the DOS near the Fermi level is larger for minority spin than majority spin in MnAs films and *vice versa* in Fe films.[18] This might account for the opposite sign of $\mathbf{S}^{FPP}$ for these two materials.

Nuclear interactions also play an important role in the electron spin dynamics in GaAs,[10] manifested here as changes in the electron spin precession frequency that saturate in several minutes. This is a result of the hyperfine interaction, wherein nuclear spins act as an effective magnetic field $\mathbf{B}_n \propto \langle \mathbf{I} \rangle / g$,[10] where $\langle \mathbf{I} \rangle$ is the average nuclear spin. As the electron spin dynamics are sensitive to the total field $\mathbf{B}_{tot} = \mathbf{B} + \mathbf{B}_n$, their Larmor frequency is indicative of $\mathbf{B}_n$. In turn, large values of $\mathbf{B}_n$ can be generated through dynamic nuclear polarization (DNP), where nuclear spins become hyperpolarized along **B** by exchanging angular momentum with a non-equilibrium electron spin distribution.[16,19] Recent studies on hybrid ferromagnet/GaAs structures have shown that DNP can be generated by the ferromagnet,[6] resulting in a nuclear polarization controlled by **M**. Initial evidence also suggested that electron spin (polarized by the ferromagnet) may drive this process. If the relevant electron spin is $\mathbf{S}^{FPP}$, then DNP gives[10]

$$\mathbf{B}_n \propto -(\langle \mathbf{S}^{FPP} \rangle \cdot \mathbf{B})\mathbf{B}/|\mathbf{B}|^2, \qquad (2)$$

for sufficiently large **B** and $\langle \mathbf{I} \rangle \ll \mathbf{I}$. The negative sign comes from the negative electron g-factor in GaAs, resulting in a $\mathbf{B}_n$ that is antiparallel to $\langle \mathbf{I} \rangle$. The inner product in Eq. (2) is consistent with data from MnAs/GaAs taken with LP and CP pumps. Referring to Fig. 2(c), $B_n(\alpha)$ (open triangles) is well fit with $B_n^0 \cos(\alpha)$ (solid line) for a LP pump, where $B_n^0$ is the fit parameter. Likewise, $B_n$ exhibits a similar dependence on $\alpha$ when the pump is CP.[6]

The opposite sign of $\mathbf{S}^{FPP}$ for MnAs and Fe enables further investigation of this ferromagnetic imprinting through disentangling the influence of $\mathbf{S}^{FPP}$ and **M** on the nuclei. This is achieved most readily with a CP pump when the samples are rotated so that the [110] (easy) axis is along **B** and the pump path is normal to the sample surface [Fig. 4(a)]. In Fe/GaAs samples, Eq. (2) predicts that $\mathbf{B}_n$ opposes **M** assuming $\mathbf{S}^{FPP}$ and **M** are parallel for a CP pump, as is the case for a LP pump. A specific value of $\mathbf{B}_n$ would then exactly cancel **B** ($B_{tot} = 0$), thus quenching electron spin precession. Figure 4(b) displays a gray-scale plot of $S_x$ vs $\Delta t$ and $B$. On the field scale shown, **M** is parallel to **B** as switching occurs at relatively small fields ($B_c \sim 0.005$ T). The influence of the Fe layer is evident—spin precession completely ceases at $B \sim \pm 0.85$ T. Correspondingly, $B_{tot}$ (open circles) vanishes at $\pm 0.85$ T, yielding a direct measure of $B_n$ for a pump power of 4.5 mW. Lowering the pump power reduces the zero-crossing field as predicted by DNP (not shown).[10] Moreover, the data indicate that $\mathbf{B}_n$ is antiparallel to both $\mathbf{M}$ and $\mathbf{S}^{FPP}$ in Fe/GaAs.

In the same manner, we determine the sign of $\mathbf{B}_n$ for MnAs/GaAs. Figure 4(c) displays a plot of $B_{tot}$ vs $B$ sweeping from negative to positive fields. For all fields except $0 < B < B_c$ (~ 0.086 T), $\mathbf{M}$ and $\mathbf{B}$ are parallel and $B_{tot}$ is larger than $B$. In the excluded range, $\mathbf{B}$ and $\mathbf{M}$ are antiparallel and $\mathbf{B}_n$ is inferred to oppose $\mathbf{B}$ due to the switching observed at $B_c$. This implies that $\mathbf{B}_n$ and $\mathbf{M}$ are parallel, and assuming that $\mathbf{M}$ and $\mathbf{S}^{FPP}$ are antiparallel for a CP pump, $\mathbf{B}_n$ must be antiparallel to $\mathbf{S}^{FPP}$. Comparing the results from Fe/GaAs and MnAs/GaAs indicates that the sign of $\mathbf{B}_n$ is not determined by $\mathbf{M}$ but rather by $\mathbf{S}^{FPP}$, in accordance with Eq. (2). Therefore, these results provide strong evidence that DNP via $\mathbf{S}^{FPP}$ is the origin of ferromagnetic imprinting of nuclear spins.

In conclusion, we have observed coherent electron spin in n-GaAs by ferromagnetic proximity polarization. Comparison of MnAs/GaAs and Fe/GaAs has given insight into this effect and the resultant imprinting of nuclei. While the origin of the effect remains unclear, the opposite sign of polarization generated by these two materials provides flexibility in orienting regions of electron and nuclear spin in a semiconductor.

We thank E. Johnston-Halperin and G. Salis for helpful discussions and acknowledge support from DARPA/ONR N00014-99-1-1096, NSF DMR-0071888, and AFOSR F49620-99-1-0033.

# References

[1] R. Fiederling et al., Nature **402**, 787 (1999). Y. Ohno et al., Nature **402**, 790 (1999).

[2] H. J. Zhu et al., Phys. Rev. Lett. **87**, 016601 (2001). A. T. Filip et al., Phys. Rev. B **62**, 9996 (2000).

[3] J. M. Kikkawa and D. D. Awschalom, Phys. Rev. Lett. **80**, 4313 (1998).

[4] J. M. Kikkawa and D. D. Awschalom, Nature **397**, 139 (1999).

[5] I. Malajovich, J. J. Berry, N. Samarth, and D. D. Awschalom, Nature **411**, 770 (2001).

[6] R. K. Kawakami et al., Science **294**, 131 (2001).

[7] S. A. Crooker et al., Phys. Rev. B **56**, 7574 (1997).

[8] M. Tanaka et al., Appl. Phys. Lett. **65**, 1964 (1994).

[9] Y. Chye et al., submitted for publication.

[10] Optical Orientation, edited by F. Meier and B.P. Zakharchenya (Elseiver, Amsterdam, 1984).

[11] J. M. Kikkawa, I. P. Smorchkova, N. Samarth, and D. D. Awschalom, Science **277**, 1284 (1997).

[12] The spin lifetime and precession frequency are similar but not identical for CP and LP pumps at $T = 5$ K, which may be attributed to the influence of (inhomogeneous) nuclear polarization.

[13] The g-factor is obtained at high temperatures (110 K), where nuclear polarization is minimal. With $B_n \sim 0$, $|g| = 0.4$ for all data in Fig. 3. The (negative) sign of g is inferred from data in Fig. 1 and Fig. 3.

[14] C. Weisbuch and C. Hermann, Phys. Rev. B **15**, 816 (1977).

[15] We do observe small phase shifts (<10°) at the switching field of **M**, which cannot be conclusively attributed to $S^{FPP}$ and may involve nuclear effects as discussed below.

[16] G. Salis et al., Phys. Rev. Lett. **86**, 2677 (2001).

[17] D. Loss, private communication.

[18] K. Shimada et al., J. Elec. Spec. **101-103**, 383 (1999). G. Chiaia, S. De Rossi, L. Mazzolari, and F. Ciccacci, Phys. Rev. B **48**, 11298 (1993).

[19] G. Lampel, Phys. Rev. Lett. **20**, 491 (1968).

**Figure Captions**

FIG. 1. Measured Faraday rotation $S_x$ vs $\Delta t$ for MnAs/GaAs using CP and LP pumps with $B$ = 0.12 T and $T$ = 5 K. Left inset: measurement geometry with angle between pump and probe (~3°) exaggerated for clarity. Right inset: normalized $S_0$ vs $E_{pump}$, with the probe energy held constant. In Figs. 1, 2, and 4, the pump and probe powers are 2.0 mW and 0.1 mW, respectively, and their energy is ~1.52 eV.

FIG. 2. Gray-scale plots of $S_x$ vs $\Delta t$ and $B$ for MnAs/GaAs, sweeping from -3 T to 3 T using LP (a) and CP (b) pump beams at $T$ = 5 K. For all gray-scale plots, white and black denote positive and negative values of $S_x$, respectively. (a) Lower panel: Fit amplitude $S_0^{LP}$ vs $B$, sweeping from - to +3 T (solid circles) and + to -3 T (open circles). (c) $B_n$ (open triangles) and normalized $S_0^{LP}$ (solid squares) vs $\alpha$ at $B$ = 0.12 T. Solid lines represent fits. $B_n$ is discussed later in the text. (c) Inset: sample geometry. Note: the MnAs/GaAs sample in (c) and Fig. 3 is different from the one used for other figures.

FIG. 3. Measurement geometry for right CP (a) and LP (b) pump beams. Measured $S_x$ vs $\Delta t$ at $T$ = 110 K for Fe/GaAs (c), (e) and MnAs/GaAs (d), (f). Solid lines represent fits. TRFR scans at $B$ = 0 were subtracted from the data in (e) and (f) to remove monotonic backgrounds. Figs. 3 (c) – (f) are plotted on the same scale. The pump and probe powers used are 4.5 and 0.45 mW, respectively, with an energy of ~1.51 eV.

FIG. 4. (a) Measurement geometry. (b) Upper panel: gray-scale plot of $S_x$ vs $\Delta t$ and $B$ for Fe/GaAs with a CP pump at $T = 5$ K. Lower panel: $B_{tot}$ vs $B$ for Fe/GaAs. (c) $B_{tot}$ vs $B$ for MnAs/GaAs (raw data not shown). Dashed lines represent $B$. The signs of $B_{tot}$ are inferred from the data. Fits to less than one TRFR oscillation have been omitted.

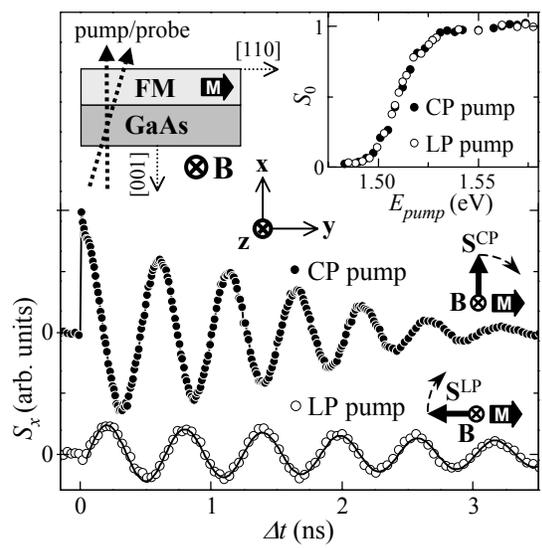

Epstein, *et al.*, Figure 1

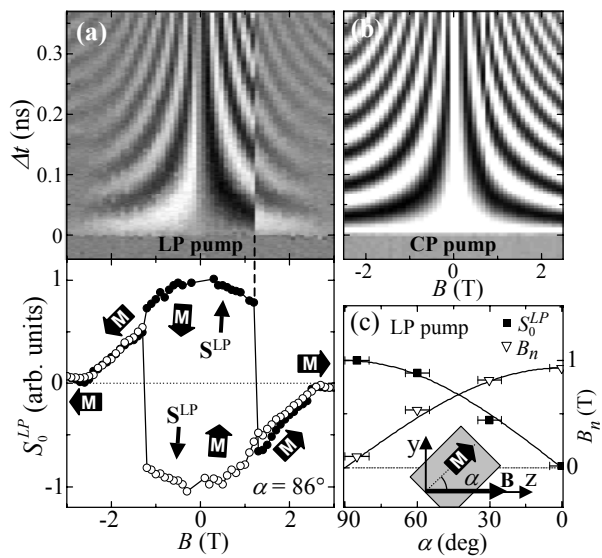

Epstein, *et al.*, Figure 2

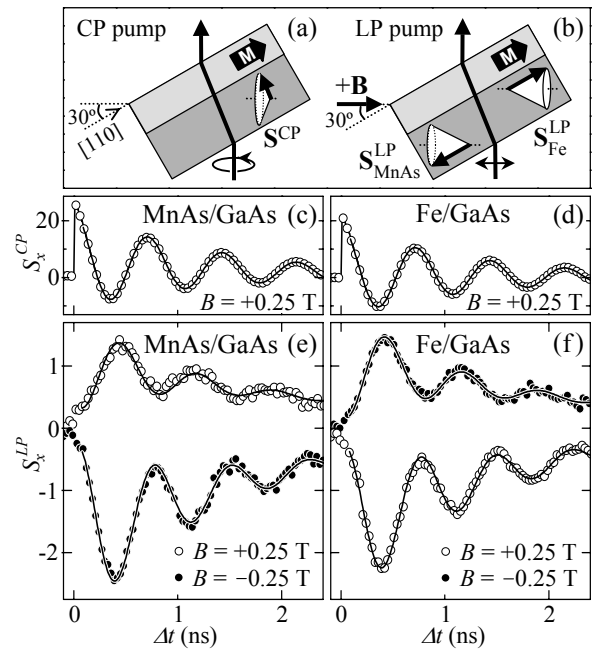

Epstein, *et al.*, Figure 3

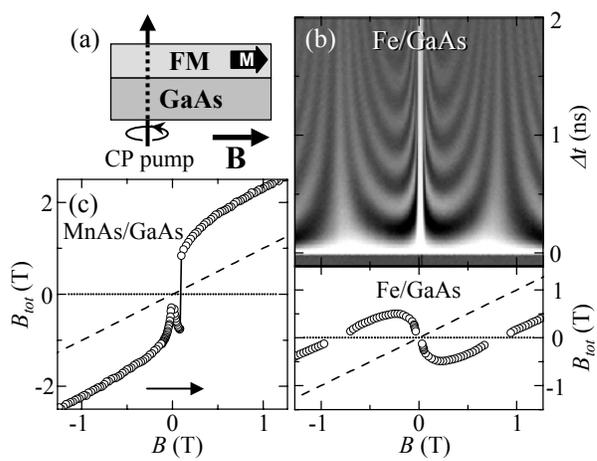

Epstein, *et al.*, Figure 4